\documentclass[11pt,fleqn]{article}
\usepackage{amssymb,latexsym}
\usepackage{amsmath, amsthm}
\newfont{\msb}{msbm10 scaled\magstep1}

\newcommand{\be}{\begin{equation}}
\newcommand{\ee}{\end{equation}}

\newcommand{\rd}{{\rm d}} 

\begin{document}
\title{\bf Unconstrained degrees of freedom for gravitational waves,
 $\beta$--foliations and spherically symmetric initial data}
\author{Jacek Jezierski\\
Albert Einstein Institut f\"ur Gravitationsphysik\\
Golm, Germany\\
and\\
Department of Mathematical Methods in Physics, \\ University of
Warsaw, ul. Ho\.za 74, 00-682 Warsaw, Poland\\
J. Kijowski \\
    Center for Theoretical  Physics, Polish Academy of Sciences\\
    al. Lotnik\'ow 32/46, 02-668 Warsaw, Poland\\}
\maketitle

\begin{abstract}
A new parameterization of unconstrained degrees of freedom for
gravitational field, used in \cite{CQG2}, has been generalized to
one-parameter family of such parameterizations, depending on a
real parameter $\beta \in [0,2]$. The description introduced in
\cite{CQG2} corresponds to the special choice $\beta=0$. The
method is closely related to the proof of the positivity of the
energy presented in \cite{PRD} where $\beta$-foliations have been
introduced (see also applications to black holes dynamics in
\cite{CQG1}, \cite{CQG2} and \cite{APP94}).

Spherically symmetric initial data corresponding to trivial
degrees of freedom is analyzed along these lines. In particular,
the quasi-local energy content of the Schwarzschild initial data
is analyzed for different choices of the $\beta$-gauge.
\end{abstract}

\section{Introduction}

We consider a compact, smooth, three-dimensional manifold $V$,
diffeomorphic to 
\[
  K(0,r_0,r_1) := \left\{ {\vec x} \in \mbox{\msb R}^3 \, \Big| \,
  (r_0)^2 \leq \sum_{i=1}^3 (x^i)^2 \leq (r_1)^2 \right\} \, .
\]
Denote by $\partial V$ its boundary. Limiting cases $r_0
\rightarrow 0$ and/or $r_1 \rightarrow \infty$ may
 be also considered.

Consider the spacetime ``tube'' $M=V\times \mbox{\msb R}^1$ and
its boundary $T=\partial V \times \mbox{\msb R}^1$ which is a
one-timelike and two-spacelike hypersurface in our spacetime.
Choose coordinates $(x^\mu)$ on $M$ in such a way that $(x^1,x^2)$
are coordinates on $\partial V$ (e.g. spherical angles $\theta$
and $\varphi$), $x^3=r$ is a ,,radial'' coordinate which is
constant on $\partial V$. Moreover, denote by $x^0$ the time
coordinate.
 So we have
  \[ V_t:=\{ x\in M \; : \; x^0=t \} =\bigcup_{r\in [r_0,r_1]} S(r)
  \quad \mbox{where} \; \; S(r):= \{ x \in V \; : \; x^3=r \} \, ,\]
\[
 T= \{ x\in M \; : \; x^3=r_0 \} \cup \{ x\in M \; : \; x^3=r_1 \} \, . \]
We use the following convention: Greek indices $\mu, \nu, \ldots$
label spacetime coordinates and run from 0 to 3; Latin indices
$k,l, \ldots$ label space coordinates on V and run from 1 to 3;
Capital indices $A,B,\ldots$ label coordinates on $\partial V$
(,,spherical angels'') and run from 1 to 2.

Consider Cauchy data $(g_{kl}, {P}^{kl})$ for Einstein equations
in the three-dimensional bounded volume $V_t$ with boundary
$\partial V_t$. This means that $g_{kl}$ is a Riemannian metric on
$V_t$ and $P^{kl}$ is a symmetric tensor density which we identify
with the ADM momentum (see \cite{ADM})
\[
{P}^{kl} = \sqrt{\det g_{mn}} ({g}^{kl} {\rm tr}\, K -  K^{kl}) \
.
\]
Here, $K^{kl}$ is the second fundamental form (external curvature)
of the imbedding of $V_t$ into the spacetime $M$ which we are
going
to construct.\\
The twelve functions $(g_{kl}, {P}^{kl})$ must fulfill four
Gauss--Codazzi constraints:
\begin{equation}
(\det g_{mn}){\cal R}- {P}^{kl}{P}_{kl}
 + \frac{1}{2} ({P}^{kl}{g}_{kl})^2  =
 16\pi(\det g_{mn}) T_{\mu \nu}n^{\mu}n^{\nu} \, ,
 \label{ws}
\end{equation}
\begin{equation}
     P_i{^l}{_{| l}} = 8\pi\sqrt{\det g_{mn}}\, T_{i\mu}n^{\mu} \, ,
        \label{ww}
\end{equation}
where $T_{\mu\nu}$ is the energy momentum tensor of the matter. By
$\cal R$ we denote the (three-dimensional) scalar curvature of
$g_{kl}$, whereas $n^\mu$ is a future timelike four-vector normal
to the hypersurface $V_t$. The geometric structure used in
(\ref{ws}) and (\ref{ww}) (the covariant derivative "$|$", rising
and lowering of the indices etc.) is the one defined by the
three-metric $g_{kl}$.

Einstein equations and the definition of the metric connection
imply the first order (in time) differential equations for
$g_{kl}$ and $P^{kl}$ (see \cite{ADM} or \cite{MTW} p. 525) and
contain the lapse function $N$ and the shift vector $N^k$ as free
parameters, canonically conjugate to the four constraints
(\ref{ws}) and (\ref{ww}):

\be \label{gdot} \dot{g}_{kl}=\frac{2N}{\sqrt{g}}\left( P_{kl}
-\frac 12 g_{kl} P \right) + N_{k|l} +N_{l|k} \ , \ee where $g:=
\det g_{mn}$ and $P:= {P}^{kl}{g}_{kl}$,
\[
\dot{P}^{kl} = -N\sqrt{g} \left( {\cal R}^{kl} -\frac 12
g^{kl}{\cal R} \right)  - \frac{2N}{\sqrt{g}}\left(
{P}^{km}{P}_{m}{^l}- \frac{1}{2} P P^{kl} \right) + \left( P^{kl}
N^m \right)_{|m} + \]
\[
+ \frac{N}{2\sqrt{g}}g^{kl} \left( {P}^{kl}{P}_{kl}-
\frac{1}{2}{P}^2 \right) -N^k{_{|m}} P^{ml}-N^l{_{|m}} P^{mk} +
\sqrt{g} \left( N^{|kl} -g^{kl} N^{|m}{_{|m}} \right)+ \]
\be\label{Pdot}
 + 8\pi N\sqrt{g} T_{mn}g^{km}g^{ln} \ .
\ee

\section{Reduced phase space of Cauchy data}
We want to analyze Cauchy data in terms of the 2+1 decomposition
of the initial surface. To describe the independent degrees of
freedom of the gravitational field we use the following objects
$(w^k,s_l)$:
\begin{equation}
 w^k:=\lambda g^{3k} \left( g^{33} \right)^{\beta-1} \, ,
 \label{wk}
\end{equation}
where $\lambda:=\sqrt{\det g_{AB}}\,$ is the two-dimensional
volume form, and
\begin{equation}
  s_A:= \left( w^3\right)^{-1} P^3{_A} \quad
   s_3 := - \left( w^3\right)^{-1} \left( \frac 12 S
  +  P^3{_A}\frac{g^{3A}}{g^{33}} \right) \, ,  \label{sk}
\end{equation}
where $\tilde{g}^{AB}$ is the two-dimensional inverse of the
two-metric $g_{AB}$, whereas by $S$ we denote $\tilde{g}^{AB}
P_{AB}$. As will be seen later, the data $(w^k,s_l)$ describe
partially reduced phase space of gravitational field.

Denote by $\sigma_{AB}$ the standard metric on a unit sphere
$S^2$, ($\sigma_{AB} \, \rd x^A \otimes \rd x^B = \rd \theta
\otimes \rd \theta + \sin^2 \theta \, \rd \varphi \otimes \rd
\varphi$) and by $\sigma=\sqrt{\det \sigma_{AB}}$($=\sin \theta$)
the corresponding volume element on the unit sphere. On each
sphere $S(r)$ we introduce the following two-metric
\begin{equation}\label{Mu}
    \mu_{AB}:=\sigma \lambda^{-1} g_{AB} \ ,
\end{equation}
conformally equivalent to $g_{AB}$. Its inverse metric is given by
$\mu^{AB}=\sigma^{-1}\lambda\tilde{g}^{AB}$.

It is easy to check that the left-hand sides of the vector
constraints (\ref{ww}) may be rewritten as follows:
\begin{equation}
P_A{^k}{_{|k}} =\left(\lambda\tilde{g}^{AC} S_{AB}\right)_{|| C}
 + (s_A w^k), {_k} - w^k s_{k,A} + \frac 12
\left(\frac{P^{33}}{g^{33}}+\beta S \right) {g^{33}} , {_A}
\end{equation}
\[
\frac{1}{g^{33}} P^{3k}{_{|k}} = \left(\frac{P^{33}}{g^{33}}
\frac{w^k}{w^3} \right), {_k} +
 \left( \frac{w^3 \tilde{g}^{AB}s_B}{g^{33}}\right), {_A}
- \lambda\tilde{g}^{AC} S_{BC} \left( \frac{w^B}{w^3} \right)_{||
A}
 + \]
\begin{equation}
+  \frac 12 S_{AB} \left(\lambda\tilde{g}^{AB}\right),{_3} \,
 + \frac{w^l}{w^3} s_l w^k , {_k} + \frac 12
\left(\frac{P^{33}}{g^{33}}+\beta S \right) \frac{w^k}{w^3} g^{33}
, {_k}
\end{equation}
where $S_{AB}:=\lambda^{-1} (P_{AB} -\frac 12 S )$ and ``$||$''
denotes two-dimensional covariant derivative with respect to the
two-metric $\mu_{AB}$.

To describe effectively\footnote{The procedure proposed here does
not cover the entire phase space but only an open neighbourhood of
the flat initial data.} the reduced phase space, i.e. the space of
classes of gauge equivalent pairs $(g_{kl}, P^{kl})$, one is free
to impose four gauge conditions which enable us to pick up a
single representative within each gauge-equivalence class.

We propose  the following  conditions:
\begin{eqnarray}
 \frac{P^{33}}{g^{33}} + \beta S &=& 0  \ , \label{gauge03S}\\
   \partial_k \left( \frac{w^k}{r^2}\right)  &=& 0 \ ,
   \label{wkk} \\
  \mu_{AB}&=&\sigma_{AB} \ . \label{gauge12}
\end{eqnarray}
It is easy to check that the gauge condition (\ref{wkk}) may be
rewritten in the following form:
\begin{equation}\label{A}
     \frac{k}{\sqrt{g^{33}}}  = \beta \frac{g^{3k}}{g^{33}}
   (\ln{g^{33}}),{_k}  -\frac{2}{x^3} \, ,
\end{equation}
where $k$ is the two-dimensional trace of the extrinsic curvature
$k_{AB}$ of $S(r)$ with respect to the three--metric $g_{kl}$ on
$V_{t}$.

Conditions (\ref{gauge03S}) and (\ref{wkk}) describe a specific
``$2 + 2$'' decomposition of spacetime. The two-parameter family
of surfaces $t=x^0=\mbox{const.}, \; r=x^3=\mbox{const.}$
(topological spheres) is defined in terms of a nonlinear system of
partial differential equations\footnote{They are quite close to
the gauge conditions discussed in \cite{EMMMWS}.} imposed on
coordinates $r$ and $t$. More precisely, eq. (\ref{gauge03S})
rewritten in terms of extrinsic curvature $K^{kl}$:
\begin{equation*}
   {\rm tr}\, K = \frac{1-\beta}{1+\beta}\frac{K^{33}}{g^{33}}
\end{equation*}
leads to the following PDE for the unknown functions $t$ and $r$:
\begin{equation} 
  \nabla_\mu \left( \frac{\nabla^\mu t}{\sqrt{(-\rd t|\rd t)}}\right)
  = \frac{1-\beta}{1+\beta}
   \nabla_\mu \left( \frac{\nabla_\nu t}{\sqrt{(-\rd t|\rd t)}}\right)
  \nabla^\mu r \frac{(\rd r|\rd t)\nabla^\nu t
  -(\rd t|\rd t)\nabla^\nu r }{(\rd r|\rd t)^2
 -(\rd t|\rd t)(\rd r|\rd r)}  \, ,
\end{equation}
where here the notation is four-dimensional with respect to the
four-metric ${\bf g}_{\mu\nu}$, e.g. $(\rd r|\rd t):= {\bf
g}^{\mu\nu}\partial_\mu r \partial_\nu t$. Similarly, the gauge
condition (\ref{wkk}) takes the following form:
\begin{equation}
\nabla_\mu \left\{{\left[ (\rd r|\rd t)^2
 -(\rd t|\rd t)(\rd r|\rd r)\right]^{\beta-\frac12}\over {r^2}
 (-\rd t|\rd t)^\beta } \left[ \nabla^\mu r
 -\frac{(\rd r|\rd t)}{(\rd t|\rd t)} \nabla^\mu t\right] \right\} = 0 \, .
\end{equation}
 However, for $\beta=1$ the
construction splits into separate equations, the first one gives a
maximal three-surface $t=\mbox{const.}$ and the second one
corresponds to a certain spherical foliation of this maximal
surface (conformally harmonic gauge in \cite{PRD}).

Condition (\ref{gauge12}) corresponds to the choice of appropriate
conformal coordinates on each two-di\-men\-sio\-nal sphere $S(r)$.
This is possible because every two-dimensional topological sphere
is conformally equivalent to a unit sphere and eq. (\ref{gauge12})
describes precisely this equivalence. However, coordinates $(x^A)$
are not fixed uniquely by the above condition but only up to the
six-parameter family of conformal transformations of the unit
sphere. We have, therefore, the residual, six-dimensional gauge
freedom on each $S(r)$. Using this freedom we may annihilate the
six-dimensional dipole component of the vector $w^k$.

Due to gauge conditions (\ref{wkk}) and (\ref{gauge12}) we can
simplify vector constraints as follows:
\begin{equation} \label{wwA}
\sigma S_A{^B}{_{|| B}} + (s_A w^k), {_k} - w^k s_{k,A} = 8\pi j_A
\, ,
\end{equation}
where $\displaystyle j_A := \sqrt{\det g_{mn}}\, T_{A\mu}n^{\mu}
$, and
\begin{equation} \label{ww3}
2\beta \left( \frac{s_m w^m w^k}{w^3} \right), {_k}
  + \left[ \sigma s^A \left( g^{33} \right)^{\beta-1} \right], {_A}
- \sigma S^A{_B} \left( \frac{w^B}{w^3} \right)_{|| A} +
 \frac{2w^l}{r} s_l  = 8\pi j_3 \, ,
\end{equation}
where $ \displaystyle j_3 := \sqrt{\det g_{mn}}\, T_{k\mu}n^{\mu}
\frac{g^{3k}}{g^{33}}$. Observe that for $\beta=1$ the ``conformal
factor'' $g^{33}$ does not enter into the equation.

Equation (\ref{wwA}) is a two-dimensional internal equation on
each slice $S(r)$ separately and it enables one to reconstruct
$S_{AB}$ from the reduced data $(s_k, w^l)$. There are, however,
additional constraints which must be fulfilled by this data. These
constraints are visible if we contract (\ref{wwA}) with an
arbitrary generator $\xi^A$ of the six-dimensional conformal group
on $S(r)$. Such a generator fulfills the conformal Killing
equation
\[
 \xi_{A || B} + \xi_{B || A} = \sigma_{AB}\, \xi^C{_{||C}} \, ,
\]
and, whence,
\[
  \sigma S_A{^B}{_{|| B}}\, \xi^A = \left( \sigma S_A{^B}\, \xi^A
  \right)_{|| B} = \partial_B \left( \sigma S_A{^B}\, \xi^A
  \right) \ .
\]
The integral of this expression over $S(r)$ vanishes identically.
Consequently, we have six residual constraints imposed on the
reduced data $(s_k, w^l)$ on each sphere $S(r)$:
\begin{equation}\label{residual}
    \int_{S(r)} \xi^A\left[ (s_A w^k), {_k} - w^k s_{k,A}\right] =
    8\pi \int_{S(r)} \xi^A j_A \, .
\end{equation}
They enable us to calculate the (six-dimensional) dipole part of
$s_A$, canonically conjugate to the dipole part of $w^A$, which
was annihilated by the residual gauge condition.

Equation (\ref{ww3}) provides a relation between variables $s_k$,
which are no longer independent parameters. If we know $s_A$ and
$w^3 \neq 0$ than (\ref{ww3}) can be viewed as an equation for the
unknown function $s_3$ with given $s_A$. Similarly, the gauge
condition (\ref{wkk}) enables us to calculate $w^3$ once we know
$w^A$.

We conclude that, similarly as in \cite{CQG2}, the (partially)
reduced canonical data $(w^k, s_l)$, fulfilling the gauge
condition (\ref{wkk}) and the residual gauge condition (the dipole
part of $w^A$ vanishes) contain all the information about the
complete Cauchy data $(P^{kl}, g_{kl})$ satisfying the constraint
eqs. (\ref{ws}), (\ref{ww})  and the gauge conditions (\ref{wkk}),
(\ref{gauge12}). In particular, to calculate the ``conformal
factor'' $g^{33}$ we must solve the scalar constraint (\ref{ws}),
which can be rewritten as follows  (see also \cite{CQG2}):
\[
2(\frac{\lambda w^k}{w^3} k), {_k} -2 \partial_A \left[ \lambda
\tilde{g}^{AB} \left(  \frac 1{\sqrt{g^{33}}} \right), {_B}
\right] +
 \frac{\lambda}{\sqrt{g^{33}}} \left( R + \frac 12 k^2 \right) = \]
\begin{equation} \label{wsred}
=\frac{\sqrt{g^{33}}}{\lambda}\left( P_{kl}P^{kl} -\frac 12 P^2
\right)
 +  \frac{\lambda}{\sqrt{g^{33}}}\left( k_{AB}
k^{AB} -\frac 12 k^2  + 16\pi\rho \right) \, ,
\end{equation}
where $\rho:=T_{\mu\nu}n^\mu n^{\nu}$ is a matter density,
$k_{AB}$ is an extrinsic curvature of two-surface $S(r)$ and $R$
is a scalar curvature of induced two-metric $g_{AB}$ (see
\cite{PRD}, \cite{CQG1} and \cite{CQG2}).

Let us notice that the following identity holds:
\[ P_{kl}P^{kl} -\frac 12 P^2 = \frac 12 \left( {P^{33} \over g^{33}}
\right)^2 + \frac 2{g^{33}}\tilde{g}^{AB}{P^3}{_A} {P^3}{_B} +
\lambda\tilde{g}^{AC} \lambda\tilde{g}^{DB} S_{AB}S_{CD} - {P^{33}
\over g^{33}} S = \] \be \label{P2} = \frac 2{g^{33}}(w^3)^2
\tilde{g}^{AB}s_A s_B + \lambda\tilde{g}^{AC}
\lambda\tilde{g}^{DB} S_{AB}S_{CD} + 2\beta(\beta+2) ( w^k s_k)^2
\, ,
 \ee 
where the last equality is implied by the gauge condition
$\displaystyle \frac{P^{33}}{g^{33}}+\beta S=0$. Let us observe
that (\ref{P2}) is nonnegative for $\beta \geq 0$. Moreover, due
to gauge condition (\ref{wkk}) and identity (\ref{P2}), the scalar
constraint (\ref{wsred}) takes the following form:
\[
 2\left[ \left( \sqrt{g^{33}}k +\frac{g^{33}}{r} \right)
 \frac{\lambda w^k}{w^3} \right], {_k}
 +\lambda R + \left[ \lambda\tilde{g}^{AB} ( \log g^{33} ), {_B}
\right],{_A} = \frac 2{\lambda}(w^3)^2 \tilde{g}^{AB}s_A s_B +
 \]
\[  +\frac{g^{33}}{\lambda}\left(
\lambda\tilde{g}^{AC} \lambda\tilde{g}^{DB} S_{AB}S_{CD} +
2\beta(\beta+2) ( w^k s_k)^2  \right)+
\frac{\lambda}{2g^{33}}\beta(2-\beta) \left(
\frac{w^k}{w^3}g^{33},{_k} \right)^2  + \] \be \label{wsf}
 +\lambda \left( k_{AB}k^{AB} -\frac 12 k^2 + \frac 12 \tilde{g}^{AB}
 ( \log g^{33} ), {_A} (\log g^{33} ), {_B}  + 16\pi\rho \right) \, .\ee
It is easy to check that for $\beta\in [0,2]$ the right-hand side
is nonnegative.  This observation enabled us to show the
positivity of ADM mass (see \cite{PRD}). For this purpose we
observed that the integral of the left hand side over the whole
$K(0,r_0,r_1)$ for  $r_0 \rightarrow 0$ and $r_1 \rightarrow
\infty$ gives the surface term at infinity, proportional to the
ADM mass.

\section{Symplectic structure and the complete reduction}

Let us observe the following identity:
\begin{equation} \label{Pdg} - P^{kl}\rd g_{kl}= 2s_k\rd w^k
+\sigma \,S_{AB} \, \rd  \mu_{AB}  + \left( \frac{P^{33}}{g^{33}}
+ \beta S\right) \rd \ln g^{33} \, ,
\end{equation}
which may be easily checked by direct computation, using
definitions (\ref{wk}), (\ref{sk}) and (\ref{Mu}). It leads to the
partial reduction of the phase space if we impose gauge conditions
(\ref{gauge03S}) and (\ref{gauge12}). More precisely, the last two
terms drop out and we are left with
\[ - P^{kl}\rd g_{kl}= 2s_k\rd w^k \, . \]
Defining
\[
  \frac {w^k}{r^2} = D^k \ ,
\]
and
\[
  2{r^2} \cdot {s_k} = A_k \ ,
\]
we get the following symplectic structure:
\begin{equation}\label{electro}
    - P^{kl}\rd g_{kl} = A_k \rd D^k \ ,
\end{equation}
together with the constraint (\ref{wkk}), which now reads:
\[
  \partial_k D^k = 0 \ .
\]
This structure is formally equivalent to the structure of
classical electrodynamics, where $D^k$ is the electric
displacement vector-density and $A_k$ is the vector potential for
the magnetic field. Constraint (\ref{ww3}) plays a role of the
non-linear ``Coulomb gauge condition'' in electrodynamics. The
analogy is not complete, because here we have also the residual
gauge condition (dipole part of $D^k$ vanishes), dual to the
residual constraints (\ref{residual}).

Further reduction of the phase space $(D^k, A_l)$ may be performed
if we observe that only two variables among $D^k$ and two
variables among $A_l$ are independent. Equations (\ref{wkk}) and
(\ref{ww3}), together with the scalar constraint (\ref{ws}),
define a subspace in the space of variables $(D^k, A_l)$),
corresponding to the two degrees of freedom of the gravitational
field. To parameterize this space by independent variables, we may
represent
\[
 A_l = {\tilde A}_l + \partial_l \phi \ ,
\]
and impose a further gauge condition:
\[
  {\tilde A}_B = \varepsilon_{BC}\ \mu^{CD} \partial_D \varphi \ .
\]
Inserting this {\em Ansatz} into (\ref{electro}) and integrating
over the entire $V$, we obtain the following symplectic form:
\[
  \Omega = \int_V \left( \rd \, D^3 \wedge \rd {\tilde A}_3
  + \rd \ (\varepsilon^{BC} \partial_B D_C) \wedge
   \rd \, \varphi \right) \ .
\]
It is obvious that the above four functions contain the entire
information about $(D^k,{\tilde A}_l)$. To reconstruct the
physical data $(P^{kl}, g_{kl})$ we must solve constraint
equations. In particular, eq. (\ref{ww3}) is a three-dimensional,
elliptic equation for the function $\phi$.

\section{Spherically symmetric initial data}
Let us assume that our initial data are spherically symmetric and
the gauge condition is compatible with spherical foliation. This
simply means that $s_A=w_A=0$ and, moreover, $w^3=r^2\sigma$ which
is consistent with (\ref{wkk}). The energy-momentum tensor is no
longer free, we have that $j_A=0$, $T_{3A}=0$ and, moreover, that
the traceless part of $T_{AB}$ vanishes:
\[ T_{AB} -\frac 12 g_{AB} \tilde{g}^{CD}T_{CD} =0 \, ,  \]
The vector constraints (\ref{wwA}), (\ref{ww3}) simplify
drastically:
\begin{equation} \label{swA}
\sigma S_A{^B}{_{|| B}} - w^3 s_{3,A} =  0 \, ,
\end{equation}
\begin{equation} \label{sw3}
2\beta \left( {s_3 w^3} \right), {_3} +
 \frac{2w^3}{r} s_3  = j_3 \, .
\end{equation}
 Assuming that $\partial_A j_3=0$ (spherical symmetry of $j_3$) we deduce
 from (\ref{sw3}) that $\partial_A s_3=0$ and from (\ref{swA}) we get
$S_{AB}=0$.

The scalar constraint is also very simple:
\[
  2\left[ \left( \sqrt{g^{33}}k +\frac{g^{33}}{r} \right)
 \lambda + r\sigma \right], {_3}  =
\]
\be \label{sws} =  2\beta(\beta+2) \frac{g^{33}}{\lambda}( w^3
s_3)^2 + \frac{\lambda}{2g^{33}}\beta(2-\beta) \left( g^{33},{_3}
\right)^2 + 16 \pi \lambda \rho \, . \ee

\subsection{Special cases $\beta=0,\frac 12, 1$ and the
Schwarzschild initial data} We would like to compare different
expressions for the Schwarzschild metric, which we obtain for
different $\beta$-foliations.

\subsubsection{Conformal gauge $\beta=1$}
Let us start with conformally flat representation of the
Schwarzschild initial data three-metric:
\begin{equation}
  \label{BH}\rd s^2 =
  \left(1+\frac{m}{2r}\right)^4 \left( \rd r^2
  +r^2\sigma_{AB}\rd x^A \rd x^B\right) \, , \quad \sigma
  = \sqrt{\det\sigma_{AB}} \; .
\end{equation}
Obviously $r=0$ corresponds to the second spatial end. One can
easily check that
\[ w^3=r^2\sigma \, , \quad \sqrt{g^{33}}=
  \left(1+\frac{m}{2r}\right)^{-2} \, , \quad
  -\frac{k}{\sqrt{g^{33}}}=\frac{2(r-\frac m2)}{r(r+\frac m2)}
\]
and
\[
 {\cal S}:= 2\left[ \left( \sqrt{g^{33}}k +\frac{g^{33}}{r} \right)
 \lambda + r\sigma \right]  = 4m\sigma\frac{r}{r+\frac m2} =
 \left\{ \begin{array}{lcr} 2m\sigma & \mbox{for} & r=\frac m2 \\
 4m\sigma & \mbox{for} & r=\infty \end{array} \right. \; .
\]
 Let us observe that the surface integral
\be
  \label{qlm} \frac1{16\pi}\int_{S(r)} {\cal S} \, ,
\ee which we obtain when integrating the scalar constraint
(\ref{sws}) over $V$, gives half of the ADM mass when calculated
on the minimal surface $S(r=\frac m2 )$ and the entire ADM mass
for $S(r=\infty )$. This means that the ``energy density'' --- the
right-hand side of the scalar constraint (\ref{sws}) --- splits
into a half which is contained outside the horizon and another
half, hidden inside the horizon. The ADM mass seen at both space
ends is equal to each other.

\subsubsection{Harmonic gauge $\beta=1/2$}
Let us denote by $R=x^3$ the solution of eq. (\ref{wkk}) for
$\beta = 1/2$, to avoid confusion with conformal coordinate $r$
introduced in the previous subsection. It is easy to check that
$R=r+\frac{m}2$. This implies the following:
\[
 2\left[ \left( \sqrt{g^{33}}k +\frac{g^{33}}{R} \right)
 \lambda + R\sigma \right]  = 4m\sigma\frac{r+\frac m4}{r+\frac m2} =
 \left\{ \begin{array}{lcr} 3m\sigma & \mbox{for} & r=\frac m2 \\
 4m\sigma & \mbox{for} & r=\infty \end{array} \right. \, .
\]
Hence, the surface integral (\ref{qlm}) gives $3/4$ of the ADM
mass when calculated on the minimal surface (and, again, the
entire ADM mass for $S(r=\infty )$).

\subsubsection{Inverse mean curvature flow $\beta=0$}
This particular case has been already analyzed in \cite{CQG2} and
\cite{APP94}. The gauge condition (\ref{wkk}) or, equivalently,
equation (\ref{A}), corresponds to the so-called inverse mean
curvature flow (see \cite{GH-TI}) and its solution enables one to
prove the Penrose inequality (see \cite{PenIneq} and references
therein). In this case equation (\ref{wkk}) implies that the
coordinate $\displaystyle x^3=r\left(1+\frac{m}{2r}\right)^{2}$ is
the usual Schwarzschild coordinate. Moreover, for the
Schwarzschild initial data and $\beta=0$ the surface integral
(\ref{qlm}) is constant on each sphere $S(r)$ and plays a role of
a {\em quasi-local mass}.

\section{Conclusions}
We hope that the description of unconstrained initial data for
gravity presented in this short article may be useful for such
applications like:
\begin{itemize}
\item construction of initial data for numerical analysis (see
e.g. \cite{BRBB}),
 \item description of the so-called dynamical
horizons (see e.g. \cite{AABK}),
 \item description of initial data
which are sufficiently close to spherical symmetry (one can use
analysis based on different $\beta$-foliations, analogous to the
one presented in \cite{CQG2} for $\beta = 0$ foliation),
 \item
construction of trapped surfaces (see e.g. \cite{MM}),
\end{itemize}
and in other cases which are not yet discovered or not known to
the authors.

\section{Acknowledgements}
This research was supported in part by the Polish Research Council
grant KBN 2 P03B 073 24 and by the Erwin Schr\"odinger Institute.


\begin{thebibliography}{66}
\bibitem{ADM} R. Arnowitt, S. Deser, C. Misner, {\it The dynamics of general
relativity}, in: Gravitation: an introduction to current research,
ed. L. Witten, Wiley, New York (1962)

\bibitem{MTW} C. Misner, K.S. Thorne, J.A. Wheeler {\it Gravitation},
W.H. Freeman and Co., San Francisco (1973)

\bibitem{PRD} J. Jezierski, J. Kijowski, Phys. Rev. D {\bf 36}, 1041-1044
(1987)

\bibitem{CQG1} J. Jezierski, Classical and Quantum Gravity {\bf 6}, 1535-1539
(1989)

\bibitem{GRG} J. Jezierski, J. Kijowski, General Relativity and
Gravitation {\bf 22}, 1283-1307 (1990)

\bibitem{CQG2} J. Jezierski,
{\it Stability of Reissner--Nordstr\"{o}m solution with respect to
small perturbations of initial data}, Classical and Quantum
Gravity {\bf 11},  1055-1068 (1994)

\bibitem{APP94} J. Jezierski,
{\it Perturbation of initial data for spherically symmetric
charged black hole and Penrose conjecture}, Acta Physica Polonica
B {\bf 25},   1413-17 (1994)

\bibitem{MM} U. Brauer, E. Malec, N. O'Murchadha,
{\it Trapped surfaces in spherical expanding open universes},
Phys. Rev. D {\bf 49}, 5601-5603 (1994);
 E. Malec, N. O'Murchadha,
{\it Trapped surfaces and the Penrose inequality in spherically
symmetric geometries}, Phys. Rev. D {\bf 49}, 6931-6934 (1994)

\bibitem{GH-TI} G. Huisken, T. Ilmanen, Journ. Diff. Geometry {\bf 59},
  353-437 (2001)

\bibitem{EMMMWS} E. Malec, M. Mars, W. Simon,
{\it On the Penrose Inequality for general horizons}
 Phys. Rev. Lett. {\bf 88},  121102 (2002)

\bibitem{PenIneq} H.L. Bray, P.T. Chru\'sciel, {\it The Penrose Inequality},
 ESI-preprint-1390, (28 pages) (Dec 2003), e-Print Archive: gr-qc/0312047

\bibitem{AABK} A. Ashtekar, B. Krishnan, {\it Dynamical horizons and their properties},
 Phys. Rev. D {\bf 68}, 104030  (25 pages) (2003)

\bibitem{BRBB} B. Reimann, B. Br\"{u}gmann,
{\it Maximal slicing for puncture evolutions of Schwarzschild and
Reissner-Nordstr\"{o}m black holes}, Phys. Rev. D {\bf 69}, 044006
(21 pages) (2004)
\end{thebibliography}
\end{document}